\documentclass[12pt,preprint]{aastex}

\def\lea{\mathrel{<\kern-1.0em\lower0.9ex\hbox{$\sim$}}}
\def\gea{\mathrel{>\kern-1.0em\lower0.9ex\hbox{$\sim$}}}

\slugcomment{Accepted by The Astronomical Journal}

\shorttitle{New Star Clusters in M33}
\shortauthors{Sarajedini et al.}

\received{2006 August 1}
\begin{document}


\title{Newly Identified Star Clusters in M33. I.\thanks{
Based on observations with the NASA/ESA {\it Hubble Space Telescope},
obtained at the Space Telescope Science Institute, which is operated
by AURA, Inc., under NASA contract NAS 5-26555. These observations are
associated with proposal GO-9873 (PI : Sarajedini)}\\  Integrated Photometry and
Color-Magnitude Diagrams}

\author{Ata Sarajedini and M. K. Barker}
\affil{Department of Astronomy, University of Florida, Gainesville, FL 32611}

\author{Doug Geisler}
\affil{Grupo de Astronomia, Departamento de Fisica, Universidad de 
Concepci\'{o}n, Casilla 160-C, Concepci\'{o}n, Chile}

\author{Paul Harding}
\affil{Astronomy Department, Case Western Reserve University, 10900 
Euclid Avenue, Cleveland, OH 44106}

\author{Robert Schommer\footnote{deceased}}
\affil{Cerro Tololo Inter-American Observatory, National Optical Astronomy 
Observatories, Casilla 603, La Serena, Chile}



\begin{abstract}
We present integrated photometry and color-magnitude diagrams (CMDs)
for 24 star clusters in M33, of which 12 were previously
uncataloged. These results are based on Advanced Camera for Surveys
observations from the Hubble Space Telescope of two fields in M33. 
Our integrated V magnitudes and V--I colors
for the previously identified objects are in good agreement with previously
published photometry.
We are able to estimate ages for 21 of these clusters using features
in the CMDs, including isochrone fitting to the main sequence turnoffs
for 17 of the clusters. Comparisons of these ages with the clusters' integrated
colors and magnitudes suggest that simple stellar population
models perform reasonably well in predicting these properties.
\end{abstract}




\section{Introduction}

Star clusters represent distinct points in space and time that are 
powerful probes of the formation and evolution of their parent galaxy. As such, it is important
to compile complete catalogs of clusters in external galaxies. After candidate
clusters are identified, followup observations are used
to confirm the membership of each cluster and determine its properties
such as metallicity (Sarajedini et al. 1998: Ma et al. 2002a), 
age (Sarajedini et al. 2000; Ma et al. 2001), 
kinematics (Schommer et al. 1991; Chandar et al. 2002), and 
structural parameters (Larsen et al. 2002). 

In the case of the late-type spiral M33, its cluster system has been studied for almost 
50 years. Beginning with the first catalog constructed by Hiltner (1960), there have
been 8 distinct compilations of cluster positions; in addition  to Hiltner
(1960), these include Melnick \& D'Odorico (1978), 
Christian \& Schommer (1982; 1988, hereafter CS), Mochejska et al. (1998), 
Chandar, Bianchi, \& Ford (1999, hereafter CBF99), 
Chandar, Bianchi, \& Ford (2001, hereafter CBF01), and Bedin et al. (2005). The first
four studies are based on ground-based observations, while the last
four make use of Hubble Space Telescope (HST) imaging data, more
specifically, the Wide Field Planetary Camera 2 and the Advanced Camera
for Surveys (ACS).

On ground-based
images with a seeing of 1 arcsec FWHM for point sources, star clusters at
the distance of M33 appear marginally resolved with typical FWHM
values of a few arcseconds. Without radial velocity information,
these cluster candidates can easily be confused with background galaxies.
In contrast, when considering HST imaging in the optical passbands, the
vast majority of star clusters are resolved into 
individual stars making their identification as genuine clusters 
essentially unequivocal.  

In the spirit of these previous studies, we present the results of a cluster
search in two ACS fields in M33. Unlike the previous studies noted above, we
have photometry for stars in each cluster so that we are able to construct 
color-magnitude diagrams (CMDs) for the clusters in our sample. This allows us to investigate 
their main sequence turnoff ages and thus any correlation between cluster age
and integrated photometric properties. The next section describes the observations and
data reduction, while Sec. 3 discusses how we identify and photometer the
clusters. The CMDs are presented and analyzed in Sec. 4 within the context of
the integrated photometry. Finally, our conclusions are given in Sec. 5.

\section{Observations and Data Reduction}

The imaging data used in this study were obtained with the Advanced Camera for
Surveys (ACS) on the Hubble Space Telescope as part of our GO-9873 program. 
The pixel scale (0.049 "/pixel) and field-of-view (3.3 x 3.3 arcmin) of ACS are ideal
for resolving clusters into individual stars for systems located in the local volume.
Two fields in M33 were observed - one in the vicinity of the star cluster M9 and 
the other including the clusters U49 and H10. Figure 1 shows the locations of
these fields while Table 1 lists the log of the observations.

A detailed description of the observations and the measurement of the
stellar photometry
has been presented by Sarajedini et al. (2006). Briefly, the pipeline-
processed FLT images were downloaded from the HST archive and
multiplied by the geometric correction image; the bad pixel masks
were applied and the resultant frames of each ACS chip (WFC1 and WFC2)
were photometered with the DAOPHOT/ALLSTAR/ALLFRAME routines
(Stetson 1994). We followed the standard procedure for producing crowded-field 
photometry using these programs (Sarajedini et al. 2000) except that we employed 
the PSFs described in Sarajedini et al. (2006).
Stars appearing on all of the frames (8 in F606W and 16 in F814W) were matched 
to form mean instrumental magnitudes. These have been corrected for the
fact that ACS suffers from problems with charge transfer efficiency (CTE) 
using the corrections of Reiss \& Mack (2004); the standardization of the photometry
has been effected using the synthetic transformations
of Sirianni et al. (2005). 

The images used in the integrated cluster photometry were produced
in the following manner. First, the pipeline-processed drizzled (DRZ) images 
for the GO-9873 program were
retrieved from the HST archive. Since we did not utilize a standard dither
pattern in obtaining the exposures, each DRZ image is composed of two
CR-SPLIT exposures at the same position. For the 4 F606W and 8 F814W DRZ 
images retrieved from the archive, we derived positional offsets in x and y to allow 
us to produce one F606W image and one F814W image using the IMSHIFT and 
IMCOMBINE tasks in IRAF. In order to estimate our photometric errors
correctly, we multiplied each DRZ image by the exposure time and added back
the background sky value that is normally subtracted as part of the pipeline
routines that produce the drizzled images. The resultant images were used
in the derivation of integrated photometry for the clusters. 

\section{Identification and Photometry}

The clusters have been identified by eye as density enhancements above the
background level. Figures 2 and 3 show an image of each cluster in the F606W filter,
where all of them have been displayed with identical grey-scaling. Table 2 lists
the position of each cluster as well as its V magnitude and V--I color.
Their (x,y) positions have been determined by convolving the combined DRZ image
in F606W with
a $\sigma$=10 pixel Gaussian kernel and then using the IRAF/IMEXAMINE task applied to
the approximate cluster centers. The resultant optimum (x,y) positions have
been transformed to Right Ascension and Declination using the World Coordinate
System in the DRZ image headers. We note that the CS cluster U49 is located on
the edge of one of the ACS field so we have excluded it from the present analysis.

The magnitude and color of each cluster have been measured using the aperture
photometry routine in DAOPHOT (Stetson 1987). To be consistent with
the previous work of CBF99 and CBF01, we have 
used an aperture radius of 2.2" to measure the magnitudes. For the
colors, CBF99 and CBF01 use a smaller aperture which is different for each cluster but
ranges from 1.0" to 2.2". We have adopted an aperture size of 1.5" for our
color measurements. The background sky is always determined in an annulus
with an inner radius of 3.5" and an outer radius of 5.0", just like CBF99 and CBF01.
The instrumental magnitudes are then corrected for the CTE effect using the equation
published by Reiss \& Mack (2004) and transformed to the standard VI system
using the synthetic equations of Sirianni et al. (2005).

Table 2 gives the resultant integrated photometry for each cluster in our sample. 
The formal errors on the magnitudes and colors are all less than 0.01 mag because
of the high S/N ratio of these clusters. 
Of the 24 clusters, half of them have been previously identified and photometered
by CBF99 and CBF01. For the V-band measurements, if we exclude 
CBF 76 and 109, which have exceptionally uncertain values as given by CBF01, we 
find a mean difference of $\langle$$\Delta$$V(Us - CBF)\rangle = 0.012 \pm 0.019$. 
For the V--I colors, we derive 
$\langle$$\Delta(V-I)\rangle = 0.058 \pm 0.025.$ Both of the quoted
uncertainties are standard errors of the mean. We see no evidence for trends
of these differences with magnitude or color as shown in Fig. 4. In fact, 
given the estimated systematic
error of $\pm$0.05 mag in the Holtzman et al. (1995) WFPC2 transformation
used by CBF99/CBF01 and a similar error in the Sirianni et al. (2005) ACS transformation,
these photometric differences are gratifyingly small. The ground-based photometry 
presented in the series of papers by Ma et al. (2001, 2002a, 2002b)
provides another check of our data. We have 12 clusters in common with this series
of papers and find a mean difference of 
$\langle$$\Delta$$V(Us - Ma)\rangle = -0.013 \pm 0.043$. Again, Fig. 4 shows
no trends with magnitude or color.

\section{Color-Magnitude Diagrams}

The integrated colors and magnitudes of the clusters are influenced by their ages,
metallicities, and reddening values. We can explore these parameters by 
constructing color-magnitude diagrams (CMDs) for each of the clusters in
our sample. Figures 5 through 28 show these CMDs; from the top and going
left to right, the CMDs represent stars within 1" of the cluster, between 1" and 2",
2" and 3", and finally 3" and 4". The solid lines are the theoretical isochrones
of Girardi et al. (2000) for ages of $10^7$, $10^8$, $10^9$, and $10^{10}$ 
years and a metal abundance of Z=0.004, which corresponds to $[M/H] = -0.7$.
This metallicity was chosen because at the mean deprojected galactocentric
distance of our two fields ($\sim$3.3 kpc), the disk abundance gradient of 
Kim et al. (2002) shows a field star abundance of $[Fe/H] $$\sim$$ -0.7$. The isochrones
have been shifted by a distance modulus of $(m-M)_0 = 24.69$ (Galleti et al.
2005) and a reddening of $E(V-I) = 0.06$ (Sarajedini et al. 2000).

Inspection of the CMDs inside 1" of the cluster centers reveals that a significant fraction 
of the clusters in
our sample exhibit clear sequences allowing us to estimate their ages. Exceptions
to this are clusters 1, 2, and 3 where the main sequence and the turnoff are not
adequately defined to allow a comparison with isochrones. In addition, as noted
by Sarajedini et al. (2000), cluster 13 (CS M9) exhibits a horizontal branch (HB)
populated both redward and blueward of the RR Lyrae instability strip suggesting
an age similar to Galactic globular clusters with similar metallicity ($[Fe/H] = -1.6$)
and HB morphology. Thus, we adopt an age of $\sim$13 Gyr for cluster 13. 
Also as noted by Sarajedini et al. (1998, 2000), the red HB clusters in M33 are
likely to be 5 to 7 Gyr younger than the genuine `old' globular clusters; as such,
we adopt an age of 6 Gyr for cluster 4 (CS H10) and cluster 10, both of which
have red HBs and main sequence turnoffs fainter than our photometric limit. Such an
old age is supported by the relatively red colors of these two clusters.

Using a value of Z=0.004, we compared each cluster CMD inside 1 arcmin with the 
Girardi et al. (2000) isochrones
primarily looking at matching the turnoff region. If needed, we adjusted the distance 
modulus and reddening under the assumption that 
$A_{F814W} = 1.85 E(F606W - F814W)$ (Sirianni et al. 2005) in order to align the main 
sequence of the isochrones with that of the data. The approximate ages and
reddenings yielded 
by this procedure are listed in Table 2. The precision of the isochrone fitting
ages is typically $\pm$0.05
in the logarithm of the age in years while the precision of the reddenings is approximately
$\pm$0.05 mag. It turns out that for ages less than $\sim$2 Gyr, the adopted metallicity 
has a negligible effect (i.e. $\sigma_{age}$$\lea$0.05 dex)
on the isochrone age for metallicities between Z=0.0004 and Z=0.019.

Ma et al. (2001, 2002b, 2004) have also measured ages for a number of M33
clusters. They used spectral energy distributions (SED) constructed with integrated 
 photometry 
in the BATC filter system (Fan et al. 1996) matched to theoretical SEDs in order
to derive the cluster ages.  For 10 clusters in common between the two studies, 
we find a mean difference of
$\langle$$\Delta$$Log Age(us - Ma)\rangle = -0.07 \pm 0.18$ with a
standard deviation of 0.55 dex. The Pearson product-moment correlation
coefficient between these values is 0.66 corresponding to a t-statistic of
2.46 which, given the degrees of freedom, yields a level of significance for
the correlation of 96\%. 

%
%
Figure 29 shows the variation of integrated cluster V--I color with the estimated
isochrone age. The top panel shows the apparent colors, the middle panel
shows the dereddened colors where a constant value of $E(V-I) = 0.06$
(Sarajedini et al. 2000) has been used for all clusters, and the lower panel 
shows the dereddened colors where the reddening from the isochrone
fits (Table 2) has been used. The lines indicate the expected variation based
on the simple stellar population models of Girardi et al. (2002) for metallicities
of Z=0.001 (dashed), Z=0.004 (solid), and Z=0.019 (dotted). 

Figure 29 reveals that
a significant fraction of the M33 clusters in our sample suffer from reddening that
is higher than what can be attributed solely to dust in the Milky Way. Therefore, it
is likely that these clusters are members of the disk of M33 where the line-of-sight
extinction is greater. When this extinction is accounted for (bottom panel), 
the scatter
of the clusters around the model lines decreases. In addition, there seems to
be general agreement between the models and the observations with younger
clusters appearing bluer in integrated light as compared with older ones.
There are a few significant outliers, but these can be plausibly attributed to
the vagaries of the observational inputs such as the reddenings and the ages. 

The behavior of the integrated absolute magnitude with age is shown in
Fig. 30 using our adopted distance modulus of  $(m-M)_0 = 24.69$. 
The dashed lines are the expected relations based on the simple
stellar population models of Girardi et al. (2002) for total cluster masses of
$10^3, 10^4, 10^5$, and $10^6 M_{\odot}$ and Z=0.004. The upper panel of Fig. 30
shows the absolute V magnitude with no correction for extinction. The middle panel 
includes a correction for extinction assuming a line-of-sight value of $E(V-I) = 0.06$ 
for all clusters. In the bottom panel, the reddening of each cluster is that required to 
facilitate the isochrone comparisons from which the ages are derived. 
The appearance of Fig. 30 suggests that the observational data are consistent
with the theoretical predictions of how simple stellar populations fade
over time. Furthermore, the tracks indicate that most of the M33 clusters in our 
sample have masses between $10^4$ and 3 x $10^4 M_{\odot}$.

\section{Summary}

In this paper, we present HST/ACS integrated photometry of 24 clusters
in M33 along with individual stellar photometry in these clusters and 
their surrounding fields. Twelve of the 24 clusters have been identified and
photometered by CBF99 and CBF01. We find mean differences of
$\langle$$\Delta$$V(us - CBF)\rangle = 0.012 \pm 0.019$ and
$\langle$$\Delta(V-I)\rangle = 0.058 \pm 0.025.$ 

Radial cuts of the photometry centered on each cluster have been performed.
For 18 of the clusters, the resulting CMDs present distinct main sequences and
turnoffs that are amenable to age determinations via  isochrone fitting. Three 
additional clusters present other CMD features that allow us to estimate
their ages. Comparison of these ages with dereddened cluster colors indicates
that the simple stellar population models of Girardi et al. (2002) perform
reasonably well in replicating this behavior. Furthermore, 
the integrated absolute magnitudes of these clusters as a function of
their ages suggests that these same models perform well in predicting the
fading clusters experience as they age. These `fading tracks' indicate
that most of the M33 clusters in our 
sample have masses between $10^4$ and 3 x $10^4 M_{\odot}$.

\acknowledgments

The authors gratefully acknowledge fruitful conversation with Rupali Chandar
as this paper was being written. Support for this work (proposal number 
GO-9873) was provided by NASA through a 
grant from the Space Telescope Science Institute 
which is operated by the Association of Universities 
for Research in Astronomy, Incorporated, under NASA contract NAS5-26555.
 D.G. gratefully acknowledges support from the Chilean
{\sl Centro de Astrof\'\i sica} FONDAP No. 15010003.

\clearpage

\begin{deluxetable}{lcccc}
\tablecaption{Observing Log}
\tablewidth{0pt}
\tablehead{
   \colhead{Field}
  &\colhead{RA  (2000)}
  &\colhead{Dec}
  &\colhead{Filter}
  &\colhead{Exp Time}
}
\startdata
U49 & 1h 33m 40s & +30$^o$ 47' 59'' & F606W & 1 x 2494s,     3 x 2640s\\
         &                       &                               & F814W & 2 x 2494s,     6 x 2640s\\
         &                       &                               &               &                  \\
M9   & 1h 34m 30s & +30$^o$ 38' 13''  & F606W & 1 x 2494s,     3 x 2640s\\
         &                       &                                & F814W & 2 x 2494s,     6 x 2640s \\
\enddata
\end{deluxetable}

\begin{deluxetable}{cccccccc}
\tablecaption{Cluster Properties}
\tablewidth{0pt}
\tabletypesize{\scriptsize}
\tablehead{
   \colhead{Identification}
  &\colhead{RA  (2000)}
  &\colhead{Dec}
  &\colhead{$V$}
  &\colhead{$V-I$}
  &\colhead{$E(V-I)$}
  &\colhead{Log t}
  &\colhead{Note\tablenotemark{a}}
}
\startdata
    &                         &                        & U49 Field              &                 &               &      \\
1  &  01:33:41.60 & +30:48:08.1 &  18.806  &  0.556  & \nodata  & \nodata &  \\
2  &  01:33:36.19 & +30:47:54.6 &  17.841  &  0.571  & \nodata  & \nodata & CBF01-109 \\
3  &  01:33:39.50 & +30:48:47.1 &  18.444  &  0.552  & \nodata  & \nodata & \\
4  &  01:33:35.13 & +30:48:59.7 &  18.360  &  1.265  & 0.25\tablenotemark{b} & 9.8  &  CS H10, CBF01-106  \\
5  &  01:33:33.29 & +30:48:30.2 &  18.416  &  0.820  & 0.30        & 8.55         &  CBF01-107 \\
6  &  01:33:34.68 & +30:48:20.9 &  19.160  &  1.080  & 0.30        &  8.75        &  CBF01-108 \\
7  &  01:33:30.91 & +30:49:11.2 &  18.538  &  0.805  & 0.45        &  8.5         &  CBF01-110 \\
8  &  01:33:32.98 & +30:49:41.1 &  18.688  &  0.975  & 0.40        & 8.5         &   CBF01-112 \\
9  &  01:33:36.78 & +30:49:17.1 &  18.657  &  0.872  & 0.45        & 7.7         &  \\
10 & 01:33:37.82 & +30:50:31.6 &  19.237  &  1.131  & \nodata  & 9.8           &  \\
11 & 01:33:45.15  & +30:49 08.8 & 19.021  &  0.695  & \nodata  & \nodata   & CBF01-64 \\
12 & 01:33:41.93  & +30:49:19.7 & 19.462  &  1.476  & 0.25        & 9.2           &  \\
      &                        &                        &  &  & \\
      &                        &                        & M9 Field  &     &  \\
13   & 01:34:30.34 & +30:38:13.4 &  17.198  &  0.978  & 0.04\tablenotemark{b} & 10.1  & CS M9, CBF01-70 \\
14   & 01:34:29.16 & +30:38:05.7 &  18.831  &  0.895  & 0.10      &  9.15      & CBF01-71 \\
15   & 01:34:33.20 & +30:38:14.6 &  19.266  &  0.626  & 0.30      &   8.7    & CBF01-76 \\
16   & 01:34:32.99 & +30:38:12.2 &  19.045  &  1.225  & 0.35      &   8.5      & CBF01-79 \\
17   & 01:34:33.18 & +30:37:36.7 &  18.246  &  0.108  & 0.20      & 7.5      & CBF01-78 \\
18   & 01:34:28.27 & +30:36:17.7 &  16.025  &  0.550  & 0.10      &  7.9      & \\
19   & 01:34:27.22 & +30:36:43.0 &  17.683  &  0.640  & 0.20      &  8.25    & \\
20   & 01:34:25.58 & +30:36:57.3 &  18.224  &  0.916  & 0.15      &  8.3      & \\
21   & 01:34:26.41 & +30:37:24.0 &  18.130  &  0.772  & 0.15      &  8.2      & \\
22   & 01:34:23.13 & +30:37:40.4 &  18.778  &  0.731  & 0.20      &  8.3    &  \\
23   & 01:34:21.68 & +30:36:46.0 &  18.391  &  0.580  & 0.10      &  8.4      &  CBF99-36\\
24   & 01:34:28.58 & +30:37:56.6 &  18.624  &  0.523  & 0.20      & 8.2       &  \\
\enddata
\tablenotetext{a}{Other identification for each cluster where CS refers
to Christian \& Schommer (1988), CBF99 refers to Chandar et al. (1999),
and CBF01 refers to Chandar et al. (2001).}
\tablenotetext{b}{Reddening value from Sarajedini et al. (2000).}
\end{deluxetable}



\clearpage
\begin{figure}[t]
\epsscale{1.0}
\caption{The location of our observed ACS fields overplotted on
the DSS-II image of M33. The field is approximately 20 arcmin a side;
North is up and east is to the left.}
\end{figure}

\begin{figure}[t]
\caption{The star clusters present in our U49 ACS field in the F606W filter.
Each image is shown with the same greyscale stretch and is
5 arcsec on a side with north up and east to the left.}
\end{figure}

\begin{figure}[t]
\caption{Same as Fig. 2 except that clusters in our M9 ACS field are shown.}
\end{figure}

\begin{figure}[t]
\caption{A comparison of our integrated cluster photometry with that of
CBF99, CBF01, and Ma et al. (2001, 2002a, 2002b). }
\end{figure}

\begin{figure}[t]
\epsscale{0.8}
\caption{Color-magnitude diagrams are shown for the region around cluster \#1
in the U49 WFC1 field. The upper left panel includes all stars within 1 arcsec of
the cluster center. The upper right panel is for stars between 1" and 2". The lower
left panel displays all stars between 2" and 3" while the lower right panel is
the CMD for all stars between 3" and 4". The isochrones are from Girardi et al.
(2000) for ages of $10^7$, $10^8$, $10^9$, and $10^{10}$ years for a metal
abundance of Z=0.004 (see text).}
\end{figure}

\clearpage
\begin{figure}[t]
\epsscale{0.8}
\caption{Same as Fig. 5 except that the CMD of cluster \#2 in the U49 WFC1 field
is shown.}
\end{figure}

\begin{figure}[t]
\epsscale{0.8}
\caption{Same as Fig. 5 except that the CMD of cluster \#3 in the U49 WFC1 field
is shown.}
\end{figure}

\begin{figure}[t]
\epsscale{0.8}
\caption{Same as Fig. 5 except that the CMD of cluster \#4 (CS H10)
in the U49 WFC2 field is shown.}
\end{figure}

\begin{figure}[t]
\epsscale{0.8}
\caption{Same as Fig. 5 except that the CMD of cluster \#5
in the U49 WFC2 field is shown.}
\end{figure}

\begin{figure}[t]
\epsscale{0.8}
\caption{Same as Fig. 5 except that the CMD of cluster \#6
in the U49 WFC2 field is shown.}
\end{figure}

\clearpage
\begin{figure}[t]
\epsscale{0.8}
\caption{Same as Fig. 5 except that the CMD of cluster \#7
in the U49 WFC2 field is shown.}
\end{figure}

\begin{figure}[t]
\epsscale{0.8}
\caption{Same as Fig. 5 except that the CMD of cluster \#8
in the U49 WFC2 field is shown.}
\end{figure}

\begin{figure}[t]
\epsscale{0.8}
\caption{Same as Fig. 5 except that the CMD of cluster \#9
in the U49 WFC2 field is shown.}
\end{figure}

\begin{figure}[t]
\epsscale{0.8}
\caption{Same as Fig. 5 except that the CMD of cluster \#10
in the U49 WFC2 field is shown.}
\end{figure}

\begin{figure}[t]
\epsscale{0.8}
\caption{Same as Fig. 5 except that the CMD of cluster \#11
in the U49 WFC1 field is shown.}
\end{figure}

\clearpage
\begin{figure}[t]
\epsscale{0.8}
\caption{Same as Fig. 5 except that the CMD of cluster \#12
in the U49 WFC1 field is shown.}
\end{figure}

\begin{figure}[t]
\epsscale{0.8}
\caption{Same as Fig. 5 except that the CMD of cluster \#13 (CS M9)
in the M9 WFC1 field is shown.}
\end{figure}

\begin{figure}[t]
\epsscale{0.8}
\caption{Same as Fig. 5 except that the CMD of cluster \#14
in the M9 WFC1 field is shown.}
\end{figure}

\begin{figure}[t]
\epsscale{0.8}
\caption{Same as Fig. 5 except that the CMD of cluster \#15
in the M9 WFC1 field is shown.}
\end{figure}

\begin{figure}[t]
\epsscale{0.8}
\caption{Same as Fig. 5 except that the CMD of cluster \#16
in the M9 WFC1 field is shown.}
\end{figure}

\clearpage
\begin{figure}[t]
\epsscale{0.8}
\caption{Same as Fig. 5 except that the CMD of cluster \#17
in the M9 WFC1 field is shown.}
\end{figure}

\begin{figure}[t]
\epsscale{0.8}
\caption{Same as Fig. 5 except that the CMD of cluster \#18
in the M9 WFC2 field is shown.}
\end{figure}

\begin{figure}[t]
\epsscale{0.8}
\caption{Same as Fig. 5 except that the CMD of cluster \#19
in the M9 WFC2 field is shown.}
\end{figure}

\begin{figure}[t]
\epsscale{0.8}
\caption{Same as Fig. 5 except that the CMD of cluster \#20
in the M9 WFC2 field is shown.}
\end{figure}

\begin{figure}[t]
\epsscale{0.8}
\caption{Same as Fig. 5 except that the CMD of cluster \#21
in the M9 WFC2 field is shown.}
\end{figure}

\clearpage
\begin{figure}[t]
\epsscale{0.8}
\caption{Same as Fig. 5 except that the CMD of cluster \#22
in the M9 WFC2 field is shown.}
\end{figure}

\begin{figure}[t]
\epsscale{0.8}
\caption{Same as Fig. 5 except that the CMD of cluster \#23
in the M9 WFC2 field is shown.}
\end{figure}

\begin{figure}[t]
\epsscale{0.8}
\caption{Same as Fig. 5 except that the CMD of cluster \#24
in the M9 WFC1 field is shown.}
\end{figure}

\begin{figure}[t]
\epsscale{0.8}
\caption{The variation of cluster integrated color with age. The upper panel shows
the apparent V--I color. The middle panel shows the reddening corrected color where
we have adopted a line-of-sight value of $E(V-I) = 0.06$ for all clusters. In the bottom
panel, the reddening of each cluster is taken from the isochrone comparison which
yields the cluster age. The solid
line is the expected relation based on a simple stellar  population with Z=0.004
from the work of Girardi et al. (2002). The dashed line is for Z=0.001 and the dotted
line is for Z=0.019.}
\end{figure}

\begin{figure}[t]
\epsscale{0.75}
\caption{The variation of cluster integrated magnitude with age adopting an absolute
distance modulus of $(m-M)_0 = 24.69$. The upper panel shows
the absolute V magnitude with no correction for extinction. The middle panel shows 
absolute V magnitude  corrected for extinction assuming a line-of-sight value of 
$E(V-I) = 0.06$ for all clusters. In the bottom
panel, the reddening of each cluster is that required to facilitate the isochrone comparisons
from which the ages are derived. The dashed
lines are the expected relations based on a simple stellar  population with Z=0.004
from the work of Girardi et al. (2002) for masses of $10^3, 10^4, 10^5$, and
$10^6 M_{\odot}$.}
\end{figure}


\begin{thebibliography}{}
\bibitem[]{} Bedin, L. R., Piotto, G., Baume, G., Momany, Y., Carraro, G., Anderson, J.,
Messineo, M., \& Ortolani, S. 2005, \aap, 444, 831
\bibitem[]{} Chandar, R., Bianchi, L., \& Ford, H. C. 1999, \apjs, 122, 431 (CBF99)
\bibitem[]{} Chandar, R., Bianchi, L., \& Ford, H. C. 2001, \aap, 366, 498 (CBF01)
\bibitem[]{} Chandar, R., Bianchi, L., Ford, H. C., \& Sarajedini, A. 2002, \apj, 564,
712
\bibitem[]{} Christian, C. A. \& Schommer, R. C. 1982, \apjs, 49, 405 (CS)
\bibitem[]{} Christian, C. A. \& Schommer, R. C. 1983, \apj, 275, 92 
\bibitem[]{} Christian, C. A. \& Schommer, R. C. 1988, \aj, 95, 704
\bibitem[]{} Fan, X. et al. 1996, \aj, 112, 628
\bibitem[]{} Galleti, S., Bellazzini, M., \& Ferraro, F. R. 2004, \aap, 423, 925
\bibitem[]{} Girardi, L., Bressan, A., Bertelli, G., \& Chiosi, C. 2000, \aaps, 141, 371
\bibitem[]{} Girardi, L., Bertelli, G., Bressan, A., Chiosi, C., Groenewegen, M. A. T.,
Marigo, P., Salasnich, B., \& Weiss, A. 2002, \aap, 391, 195
\bibitem[]{} Hiltner, W. A. 1960, \apj, 131, 161
\bibitem[]{} Kim, M., Kim, E., Lee, M. G., Sarajedini, A., \& Geisler, D. 2002, \aj,
123, 244
\bibitem[]{} Ma, J., Zhuo, X., Wu, H., Chen, J., Jiang, Z., Zhu, J., \& Xue, S.
2001, \aj, 122, 1796
\bibitem[]{} Ma, J., Zhou, X., Wu, H., Chen, J. -S., Jiang, Z. -J., Xue, S. -J.,
\& Zhu, J. 2002a, ChJAAp, 2, 127
\bibitem[]{} Ma, J., Zhou, X., Chen, J., Wu, H., Jiang, Z., Xue, S., \& Zhu, J.
2002b, \aj, 123, 3141
\bibitem[]{} Ma, J., Zhou, X., \& Chen, J. -S. 2004, ChJAAp, 4, 125
\bibitem[]{} Melnick, J., \& D'odorico, S. 1978, \aaps, 34, 249
\bibitem[]{} Mochejska, B. J., Kaluzny, J., Krockenberger, M., Sasselov, D. D.,
\& Stanek, K. 1998, AcA, 48, 455
\bibitem[]{} Reiss, A.\& Mack, J.  2004, ACS-ISR 2004-06
\bibitem[]{} Sarajedini, A., Geisler, D., Harding, P., \& Schommer, R. 1998, \apj,
508, L37
\bibitem[]{} Sarajedini, A., Geisler, D., Schommer, R., \& Harding, P. 2000, \aj, 
120, 2437
\bibitem[]{} Sarajedini, A., Barker, M. K., Geisler, D., Harding, P., \&
Schommer, R. 2006, \aj, in press
\bibitem[]{} Schommer, R. A., Christian, C. A., Caldwell, N., Bothun, G. D.,
\& Huchra, J. 1991, \aj, 101, 873
\bibitem[]{} Sirianni, M. et al. 2005, \pasp, 117, 1049
\bibitem[]{} Stetson, P. B. 1987, \pasp, 99, 191
\bibitem[]{} Stetson, P. B. 1994, \pasp, 106, 250
\end{thebibliography}
\end{document}